\documentstyle[11pt,aaspp]{article}
\input{psfig}
\eqsecnum
\def\beq{\begin{equation}}
\def\eeq{\end{equation}}
\begin{document}
\title{A Laboratory for Magnetized Accretion Disk Model: Ultraviolet 
and X-ray Emission from Cataclysmic Variable GK Persei}
\author{Insu Yi$^1$ and Scott J. Kenyon$^2$}
\affil{$^1$Institute for Advanced Study, Olden Lane, Princeton, 
NJ 08540; yi@sns.ias.edu}
\affil{$^2$Smithsonian Astrophysical Observatory, 60 Garden St., 
Cambridge, MA 02138; skenyon@cfa.harvard.edu}

\begin{abstract}

We analyze the ultraviolet spectrum of the cataclysmic variable GK Per
at maximum light.  The flat ultraviolet spectrum in this system requires a
truncated inner accretion disk and an unusually flat radial temperature
profile.  This requirement is not satisfied by any non-magnetic steady
or non-steady disk model.

We consider a magnetized accretion disk model to explain the ultraviolet 
spectrum.  The available data on the white dwarf spin and possible 
quasi-periodic oscillations constrain the magnetic field, $B_{*}$, and 
the disk accretion rate, ${\dot M}$, to lie along a well-defined 
spin-equilibrium condition $({\dot M}/10^{17} ~ \rm g ~ s^{-1}) \sim 
100(B_{*}/10^7G)^2$.  
Our self-consistent treatment of the magnetic torque on the disk flattens 
the disk temperature distribution outside the disk truncation radius.  
This modified temperature distribution is too steep to explain the 
UV spectrum for reasonable field strengths.

X-ray heating is a plausible alternative to magnetic heating in GK Per.
We estimate that the disk intercepts $\sim$ 5\% of the accretion energy
in outburst, which results in an extra disk luminosity of $\sim$ 5--10
$L_{\odot}$.  Model spectra of optically thick disks are too blue to
match observations.  The UV spectrum of an optically thick disk with an
optically thin, X-ray heated corona resembles the observed spectrum.
The X-ray luminosity observed during the outburst indicates
${\dot M}<10^{18} ~ \rm g ~ s^{-1}$, which is a factor of 10 lower 
than that required to explain the ultraviolet luminosity.  Radiation
drag on material flowing inward along the accretion column lowers the
shock temperature and reduces the X-ray luminosity.  Most of the
accretion energy is then radiated at extreme ultraviolet wavelengths.

\end{abstract}


\section{Introduction}

The old nova GK Persei (Nova Per 1901) is an intriguing cataclysmic
variable (CV).  It is one of a few old novae known to undergo dwarf 
nova outbursts, albeit with the unusually long recurrence time of
$\sim$ 400 d (Sabbadin \& Bianchini 1983).  
The system has a coherent 351 s X-ray pulsation (Norton et al. 1988, 
Norton \& Watson 1989, Ishida et al. 1992) that has also been detected
in ultraviolet light curves (Patterson 1991).  Longer wavelength optical
light curves show quasi-periodic oscillations at periods of 360--400 s
(Patterson 1991).  Although the optical spectrum of GK Per is normal
compared to other long period CVs, its ultraviolet (UV) spectrum has a 
flat continuum and strong emission lines compared to other dwarf novae.
The emission lines remain prominent throughout outburst, even though
the UV continuum temperature is cool, $\sim 10^4$ K.


GK Per has many properties in common with the DQ Her subclass of CVs, 
which are magnetic binaries also known as intermediate polars (Warner 1985). 
Most of the system's `unusual' properties are then understood if an
accretion disk fed by the lobe-filling K0 secondary is truncated by
a magnetic field well above the white dwarf photosphere.  The X-ray
pulsations result from `hot spots' where the magnetic accretion columns
impact a white dwarf with a rotation period of 351 s.  The flat UV spectrum
and the long recurrence time for dwarf nova outbursts also support this
interpretation. 
For outbursts beginning in the inner region of the accretion disk
(e.g. Smak 1991,1993, Kim et al. 1992), 
assuming that the effective disk temperature
in the inner region is constant, the recurrence time scale is roughly
the viscous time scale $\propto R^2/\nu$ where $R$ is the radius of the
thermally unstable inner region of the disk and $\nu$ is the kinematic 
viscosity coefficient (Smak 1993).
Using Smak's (1993) formula for the critical surface density, $\Sigma_{crit}$,
for the cold branch disk and $\nu\sim {\dot M}/3\pi\Sigma_{crit}$
(e.g. Frank et al. 1992), the recurrence time scale can be estimated as
\beq
t_{rec}\sim [1.4\times 10^4 s] \left(\alpha\over 0.01\right)^{-0.79}
\left(M_*\over M_{\sun}\right)^{1.74}\left({\dot M}\over 10^{15} g/s
\right)^{-1}\left(R\over 10^{9} cm\right)^{3.11}
\eeq
where $\alpha$ is the Shakura-Sunyaev viscosity parameter
(e.g., Frank et al. 1992), $M_*$ is the mass of the accreting white 
dwarf, and ${\dot M}$ is the mass accretion rate in the inner region of the
the cold disk (which is different from the mass transfer rate).
The recurrence time scale depends very weakly on the mass transfer rate
(e.g. Cannizzo \& Mattei 1992). For a typical quiescent $\alpha\sim 0.01$, 
the observed long recurrence time scale implies a truncated disk with
an inner radius of $R_0>10^{10}$ cm (Cannizzo \& Kenyon 1986, 
Angelini \& Verbunt 1989, Kim et al. 1992).  The lack of disk material 
close to the white dwarf also produces a flat UV continuous spectrum 
by reducing the characteristic disk temperature.

Despite the conceptual simplicity of this interpretation, several 
observations of GK Per remain poorly understood.  The X-ray luminosity
is a small fraction of the UV luminosity in outburst and quiescence
as in other DQ Her systems (Patterson 1994).  Both the flat UV spectrum
and the outburst recurrence time scale require inner disk radii large
compared to the corotation radius of a white dwarf with a 351 s spin 
period, $R_0>10^{10}$ cm vs $R_c \sim 8 ~ \times ~ 10^9$ cm.  
Accretion should not occur in the circumstances.  The UV spectrum
further requires larger values for both $\dot M$ and $R_0$ than
allowed by disk instability models for the eruption.

In this paper, we consider GK Per as a laboratory to test our understanding 
of the physics of the disk-magnetosphere interaction.  Besides disrupting
the disk at an inner radius larger than the stellar radius,  the stellar
magnetosphere redistributes angular momentum and energy through torques on 
disk material (Ghosh \& Lamb 1979ab).  This interaction modifies the disk 
temperature distribution and accretion luminosity in a predictable way
that observations can test directly (Mauche et al. 1990, Kenyon et al. 1996)
We focus on the dissipation in the disk with some recent improvements and 
use emission from the accretion column and the disk as diagnostic tools for 
magnetized accretion disk models.  Our goal is to use the observed UV spectrum 
and various X-ray constraints to develop a self-consistent picture for this
and other DQ Her system.  We begin with a discussion of simple disk models
in \S2, describe and apply a magnetized disk model in \S, consider X-ray 
heating of the disk in \S4, and conclude with a brief summary and discussion
in \S5.  

\section{The Observed UV Spectrum and Simple Disk Models for GK Per}

The UV spectrum of GK Per at maximum is very unusual compared 
to other dwarf novae in outburst (see also Wu et al. 1989).
The dereddened continuum is almost flat in $F_{\lambda}$ over 
1200--3200~\AA~and closely resembles the continuum of a B8-B9 V star 
with an effective temperature of 10,000--12,000 K (Wu et al. 1989).  
In contrast, the UV continua of most dwarf novae at maximum resemble 
spectra of much hotter B2-B3 V-III stars having temperatures of 
$\sim$ 20,000 K (la Dous 1991).  The UV spectrum of GK Per also 
contains strong He~II, C~IV, and N~V emission lines.  These lines
are usually strong absorption features in dwarf novae at maximum,
although some have very weak emission lines (la Dous 1991).  The
strong emission features in GK Per are similar in strength to lines
observed in other DQ Her-type stars in the high state (la Dous 1991).
The He~II line in GK Per is, however, exceptionally strong compared 
to lines observed in most DQ Her stars.

Bianchini \& Sabbadin (1983) first showed that standard steady-state 
disk models could not reproduce UV observations of GK Per at minimum.
Wu et al. (1989) reached the same conclusions for outburst spectra.  Both 
groups showed that disks with an inner hole could roughly match observed 
fluxes.  Bianchini \& Sabbadin (1983) estimated an inner disk radius of
$R_0 \sim 10^9$ cm for $\dot M \approx 2 \times 10^{16} ~ \rm g ~ s^{-1}$
at minimum; Wu et al. preferred a much larger inner radius,
$R_0 \sim 3-5 \times 10^{10}$ cm, in outburst when 
$\dot M \approx 5-20 \times 10^{19} ~ \rm g ~ s^{-1}$.
Wu et al. (1989) remarked that the UV data do not constrain $R_0$ and
$\dot M$ independently, because any disk model with a maximum temperature 
of $\sim 10^4$ K produces a flat spectrum similar to the observed spectrum.
Their set of models has $\dot M$ slowly decreasing with increasing $R_0$
to reproduce the observed spectrum.


To compare model disk spectra with observations, we first verified the 
Wu et al. (1989) measurements using {\it IUE} data at maximum light 
collected from the NSSDC archives and binned the $F_{\lambda}$ data, 
dereddened by $E_{B-V}$ = 0.3, in 25~\AA~intervals as in Wu et al.  
The filled circles in Figure 1 plot our results.  The observed fluxes 
have an intrinisic uncertainty of $\pm$10\% as indicated by the error bar.  
The probable error in the reddening, $\delta E_{B-V} \approx$ 0.05, 
introduces an additional error of $\pm$25\% at 1300~\AA~and $\pm$10\% at 
3000~\AA.  We then computed spectra for steady-state disk models in which 
each annulus radiates as a main sequence star (see Cannizzo \& Kenyon 1986).  
For the central star, we adopt $M_* = 0.9~M_{\odot}$ (cf. Wu et al. 1989, 
Watson et al. 1985, Mauche et al. 1990, Kim et al. 1992) and 
$R_*=6.22\times 10^8~\rm cm$ from the white dwarf mass-radius relation 
(Nauenberg 1972, Ritter 1985).  The disk has a corotation radius of 
$R_c=7.20 \times 10^{9} ~ \rm cm$ for a rotational period of $P_*=351$ s 
and an outer radius of $R_{out}=1.5 \times 10^{11} ~ \rm cm$ appropriate 
for the K0 IV secondary (e.g., Cannizzo \& Kenyon 1986).  We adopt a disk 
inclination of $i$ = 60$^{\circ}$ and a distance of 470 pc 
(e.g., Bianchini \& Sabbadin 1983, Crampton et al. 1986).  
The uncertainty in the absolute level of our predicted fluxes is roughly
$\pm$50\% for reasonable uncertainties in the mass, radii, inclination,
and distance. The slope of our predicted spectra have higher accuracy, 
$\pm$20\%, as long as main sequence stars are a reasonable approximation
to the emitted spectrum.  We thus consider a model successful if it 
matches the observed fluxes to a factor of 2 and the spectral slope to 
$\pm$25\%.

Figure 1 compares spectra of two steady disk models with the dereddened
UV continuum data.  We computed models for $R_0$ = 0.9 $R_c$ and 
$\dot M = 8.8 \times 10^{18} \rm ~ g ~ s^{-1}$ (SS1) and 
$R_0$ = 5 $R_c$ and $\dot M = 4 \times 10^{19} \rm ~ g ~ s^{-1}$ (SS2).
Neither model reproduces the data very well.  The first model (SS1) has 
an inner radius near the corotation radius and also has roughly the correct 
bolometric luminosity, L $\approx$ 20 $L_{\odot}$.  This model -- and any other
steady disk model with $R_0 \approx R_c$ -- is too blue due to the high 
temperatures, $\sim$ 2--3 $\times 10^4$ K, near the corotation radius.  
The second model (SS2) has the correct UV slope but is not luminous enough.  
More luminous models are hotter than this model and fail to match 
the UV slope.  We could match the UV data better with a smaller $i$ for 
model SS2, but observations of binary motion in GK Per preclude 
$i \le$ 45$^{\circ}$ (Crampton et al. 1986).

Non-steady disk models also fail to match UV observations of GK Per.
Kim et al. (1992) developed disk instability models to account for the 
system's dwarf nova eruptions and matched the visual magnitude and
recurrence time of typical eruptions.  We read the temperature distribution
for their model 82 from their Figure 8 to compute the `DI' spectrum 
shown in Figure 1.  Our calculation reproduces Kim et al.'s spectrum
in their Figure 9 to $\sim$ 10\% as judged by a visual comparison.  
The spectrum of this model resembles our second steady-state model 
and also falls short of the observations.  
Although the DI and SS2 models reproduce the observed maximum visual
magnitude of GK Per, both have $R_0 \gg R_c$.  This result is contrary 
to our expectations in a magnetically truncated disk model.  We expect
$R_0 \le R_c$, which allows disk material to match the rotational 
velocity of the central star and accrete along magnetic field lines.

The failure of standard disk instability and steady-state disk 
models suggests that another mechanism heats the disk.  To produce a flat 
UV spectrum in the disk, we require a flatter radial temperature gradient 
than in a standard steady-state disk model (e.g., $T \propto R^{-x}$, with
$x \approx$ 0.5--0.6).  Non-steady disks often have such flat temperature
gradients in {\it quiescence}, when the disk mass grows with time (see
Kim et al. 1992).  Most non-steady disk models closely approximate steady
disks with $x \approx 0.75$ during outbursts (see Kim et al. 1992 and 
references therein).
The UV spectra of currently available non-steady models are also not 
luminous enough (Figure 1).  External heating mechanisms, however, can
produce flat temperature gradients.  Ghosh \& Lamb (1979ab) first showed
that the magnetic torque at the inner edge of a magnetically truncated disk 
acts to flatten the disk temperature gradient close to the corotation radius.
Mauche et al. (1990) applied this concept to GK Per and reproduced some
aspects of the observations.  X-ray heating is another external mechanism
that can flatten the disk temperature distribution.  Although disk irradiation
from an accretion hot spot or boundary layer is unimportant in most CVs
(Wade 1988; Smak 1989), the large disk in GK Per can intercept and reradiate
a significant fraction of the luminosity emitted from the central white dwarf.
The temperature distribution of a concave, illuminated disk can approach
$T \propto R^{-1/2}$ in some disks (e.g., Kenyon \& Hartmann 1987).
We will first consider magnetic heating in \S3 and then develop a
simple X-ray heating model in \S4.

\section{Disk-Magnetosphere Interaction and Magnetic Heating}

\subsection{Magnetic Truncation Radius}

To estimate the magnitude of disk heating in the GK Per disk, we
need to derive the disk truncation radius, $R_0$, and the magnetic
torque on the disk, $N$.  To compute these quantities, we follow 
Yi (1995) and adopt a geometrically thin, optically thick accretion
disk in which the viscous angular momentum transport is parametrized 
by the conventional $\alpha$-viscosity (e.g., Frank et al. 1992;
see also Wang 1987; Campbell 1992; Kenyon, Yi, \& Hartmann 1996).
The magnetic stress truncates the disk roughly where magnetic
pressure halts the slow inward drift of disk material.  This stress
also provides an additional dissipative heating term.  Both the
truncation radius and the heating are essential ingredients for a 
self-consistent model of the observed disk emission.  Neither of
these quantities can be derived completely from first principles:
we thus follow previous investigations and parameterize the interaction
between the disk and the magnetic field.

We assume a dipolar magnetic field structure, in which the $z$-component 
of the stellar magnetic field is
\beq
B(R)\approx B_{z}(R)=-{\mu_{*}\over R^3} ~ ,
\eeq
where $\mu_{*}=B_*R_*^3$ is the magnetic moment of the star, $B_*$ is
the stellar surface field strength, and $R_*$ is the stellar radius.
The azimuthal component of the induction equation in cylindrical 
coordinates $(R,\phi,z)$ is
\beq
{\partial B_{\phi}\over \partial t}\approx B_{z}{\partial v_{\phi}\over
\partial z}-{\partial v_{z} B_{\phi}\over\partial z}
+{\partial\over\partial z}\left(\eta_{diff}{\partial
B_{\phi}\over\partial z}\right)
\eeq
where $\eta_{diff}$ is the magnetic diffusivity.
$B_{\phi}$ is amplified by differential rotation in the disk.
To match the angular velocity of disk material with the star,
we adopt a vertical gradient of the rotational velocity,
$v_{\phi}=R\Omega$, as
\beq
\left|{\partial v_{\phi}\over \partial z}\right|
\approx \gamma\left|\Omega_{*}-\Omega\right|
\eeq
where $\Omega_*$ is the stellar angular velocity, $\Omega$ is the
disk (Keplerian) angular velocity, and
$\gamma$ is a parameter which accounts for the uncertainty in the
vertical velocity shear (Wang 1987, 1995, Yi 1995). 
The amplification dominated by differential rotation is then
\beq
\left|\partial B_{\phi}\over \partial t\right|_{+}\approx \gamma\left|
\Omega_{*}-\Omega\right|\left|B_{z}\right| ~ .
\eeq
We model the diffusive loss term by assuming 
that the magnetic Prandtl number is of order unity:
$\eta_{diff}\sim\alpha$ (Yi 1995, Campbell 1992).
This expression is appropriate when turbulent diffusion is responsible 
for both the diffusive field loss and the viscous angular momentum 
transport. The diffusive loss time scale is
$\tau_{d}\approx (\alpha\Omega)^{-1}$ and 
\beq
\left|\partial B_{\phi}\over \partial t\right|_{-}={\left|B_{\phi}
\right|\over\tau_{d}}
\approx \alpha\Omega\left|B_{\phi}\right|.
\eeq
In steady state, we equate the amplification and diffusive loss terms:
\beq
\left|B_{\phi}\over B_z\right|=\left|\gamma(\Omega_{*}-\Omega)\over
\alpha\Omega\right|.
\eeq

We set the inner disruption radius, $R_0$, by balancing magnetic force
with the force of disk material drifting inward (Wang 1987, Campbell 1992, 
Yi 1995):
\beq
-[B_zB_{\phi}]_{z=H}R^2={\dot M}{d\over dR}(\Omega R^2) ~ .
\eeq
The disruption radius lies inside the corotation radius, $R_c$, where
disk material corotates with the stellar magnetosphere.  We evaluate the
field components at the disk photosphere, as indicated by the subscript 
``$z=H$.'' With this prescription, the disruption radius is computed from
\beq
\left(R_0\over R_c\right)^{7/2}=A\left|1-\left(R_0\over R_c\right)^{3/2}
\right|
\eeq
where $R_c=(GM_*P_*^2/4\pi^2)^{1/3}$ is the corotation radius and
$P_*=2\pi/\Omega_*$ is the stellar spin period.
The dimensionless constant $A$ is (e.g. Yi 1995)
\beq
A=[9.1\times 10^{4}]
B_{eff,7}^2 R_{*,9}^6 M_{*,1}^{-5/3} P_{*,2}^{-7/3}
{\dot M}_{17}^{-1}
\eeq
where $B_{eff,7}=\sqrt{\gamma/\alpha}B_*/10^7~ \rm G$, $R_{*,9}=
R_*/10^9~ \rm cm$, $M_{*,1}=M_*/M_{\odot}$, $P_{*,2}=P_*/100 ~ \rm s$, 
and ${\dot M}_{17}={\dot M}/10^{17} ~ \rm g ~ s^{-1}$.

Kenyon et al. (1996) compared the truncation radius derived from
equation (3-8) with the Ghosh \& Lamb (1979ab) model for parameters
appropriate to pre-main sequence stars.  In general, $R_0$ is approximately
equal to $R_c$ for systems with short rotational periods, low
accretion rates, and high magnetic field strengths.  The disruption
radius moves closer to the central star as $\dot M$ and $P_*$ increase,
and as $B_*$ decreases.  Compared to Ghosh \& Lamb (1979ab) models, our treatment
yields larger $R_0$ for high $\dot M$ and high $P_*$ systems; we derive
smaller $R_0$ than Ghosh \& Lamb (1979ab) when $\dot M$ is low and $P_*$ is small.

\subsection{Magnetic Torque and Disk Heating}

The interaction between the stellar field and the accretion disk produces a 
torque on both the star and the disk.  The torque on the star has three parts
(Ghosh \& Lamb 1979ab): (i) a spin-up torque from disk material with 
$R<R_c$, (ii) a spin-down torque from disk material with $R>R_c$, and 
(iii) an accretion torque carried by gas falling onto the central star.
The total torque is 
\beq
N={7\over 6} N_0 {1-(8/7)(R_0/R_c)^{3/2}\over 1-(R_0/R_c)^{3/2}}
\eeq
where $N_0={\dot M}(GM_*R_0)^{1/2}$ (e.g., Yi 1995, Campbell 1992; 
Ghosh \& Lamb 1979ab).
The equilibrium spin $N=0$ occurs when $R_0/R_c=0.915$ (see also Wang 1995).
Li (1996) derives a smaller value of $R_0/R_c = 0.7$ for equilibrium spin
from another treatment of the magnetic interaction between the star and
the disk.  This difference illustrates the uncertainty of parameterized
magnetic disk models.

To derive the disk temperature structure, we adopt the vertically averaged 
versions (Mauche et al. 1990) of the momentum equations and energy equation 
which includes the magnetic torque and dissipation (e.g. Shu 1992).
The $\phi$ component of the momentum equation is
\beq
{v_R\over R}{\partial\over\partial R}\left(R^2\Omega\right)=
{1\over R^2\rho}{\partial\over\partial R}\left(\rho\nu R^3
{\partial\Omega \over R}\right)
+{B_z\over 4\pi\rho}{\partial B_{\phi}\over \partial z}
\eeq
where $\nu$ is the kinematic viscosity coefficient. 
Mass conservation gives
\beq
{\dot M}=-2\pi R \Sigma v_R=-\int_{-H}^{H} 2\pi R\rho V_Rdz
\eeq
where $\Sigma$ is surface density of the disk and $H$ is the 
half-thickness of the disk. Combining (3-11) and (3-12), we get
\beq
-{\dot M\over 2\pi R^2}{\partial\over \partial R}(R^2\Omega)=
{1\over R^2}{\partial\over\partial R}\left(\nu\Sigma R^3{\partial\Omega
\over\partial R}\right)+{1\over 2\pi}[B_{\phi}B_z]_{z=H}.
\eeq
Defining the viscous torque
\beq
N_{visc}=-2\pi\nu\Sigma R^3{\partial\Omega\over\partial R}
\eeq
and the magnetic torque
\beq
{dN_{mag}\over dR}=-R^2[B_{\phi}B_z]_{z=H}=-R^2{\gamma\over \alpha}
{\Omega_*-\Omega\over \Omega}B_z^2
\eeq
gives the simple integrated momentum equation
\beq
N_{visc}={\dot M}(R^2\Omega-R_o^2\Omega_o)-\int_{R_o}^{R}dR^{\prime}
{dN_{mag}\over dR^{\prime}}.
\eeq

The total dissipation rate in the disk comes from energy conservation
\beq
{dL\over dR}={\dot M}{d\over dR}\left(-{GM\over R}
+{1\over 2}R^2\Omega^2\right)-{d\over dR}(\Omega N_{visc})-\Omega_*
\left(dN_{mag}\over dR\right)
\eeq
where $dL/dR$ is the differential luminosity from an annulus between
$R$ and $R+dR$. 
For a disk surface radiating as a blackbody, the effective 
disk surface temperature is $T_{eff}=[(dL/dR)/4\pi\sigma R]^{1/4}$.
Combining (3-17) with (3-16), we get
\beq
{dL\over dR}=-N_{visc}{d\Omega\over dR}-(\Omega_*-\Omega)
{dN_{mag}\over dR}
\eeq
and
\beq
{\dot M}(R^2\Omega-R_o^2\Omega_o)=N_{visc}+\int_{R_o}^RdR^{\prime}
{dN_{mag}\over dR^{\prime}}.
\eeq

To determine the temperature distribution for a specific model, we
adopt a set of fixed stellar parameters ($R_*,M_*,P_*$) and disk
parameters ($\alpha$, $\gamma$) and choose the two free parameters 
(${\dot M},B_{*}$).  Observations do not directly constrain the disk 
parameters.  The temperature distribution of the magnetized disk is 
specified by a unique combination $\sqrt{\gamma/\alpha}B_{*}$ rather 
than individual values of $B_{*},\alpha,\gamma$ (Kenyon et al. 1996) 
as long as the disk remains optically thick. We adopt $\alpha=0.3$ and
$\gamma=1$. With the physical variables set, 
we solve eq. (3-8) for the disk truncation radius, $R_0$. 
By combining $R_0$ with eqs. (3-15) and (3-19), we derive $N_{vis}$ 
and $N_{mag}$. Finally, the differential luminosity distribution $dL/dR$ 
-- and hence the effective disk temperature $T_{eff}\propto (dL/dR)^{1/4}$ 
-- is obtained from eq. (3-18).

\subsection{Emission from a Magnetized Disk}

We now apply the magnetized disk model to observations of GK Per.
Our model must first satisfy the X-ray period constraint, which
places the corotation radius at $R_c = 7.2 \times 10^9$ cm as 
described in \S2.  Accretion occurs only if the truncation radius,
$R_0$, lies inside $R_c$, which precludes disk models with $R_0 \gg R_c$.  
The observations indicate that the white dwarf is close to spin 
equilibrium, which favors models with $R_0 \sim R_c$ as described above.  
Patterson (1991) estimates a mean spin-up rate of 0.0008 sec yr$^{-1}$ 
from optical data.  The X-ray timing data are consistent with this rate, 
although a constant period can fit the X-ray data alone (Ishida et al. 1992).  
In our model, the white dwarf achieves spin equilibrium when $N = 0$ at 
$R_0/R_c=0.915$ in eq. (3-10), which implies $A=5.856$ in eq. (3-8).  
If we adopt the white dwarf parameters from \S2, eq. (3-9) requires a 
simple spin equilibrium curve:
\beq
{\dot M}_{17}=57 B_{eff,7}^2
\eeq

Patterson's (1991) spin-up rate similarly limits the accretion torque
to $|N|\le 5.9\times 10^{34}$ g cm$^2$ s$^{-2}$ for a white dwarf moment 
of inertia, $I_*=2M_*R_*^2/5=2.8\times 10^{50}$ g cm$^2$. The spin-up line 
is then 
\beq
{\dot M}_{17}=\left(A\over 8.21\times 10^2\right)^{-1} B_{eff,7}^2
\eeq
where $A$ is 
\beq
A=x^{7/2}/\left|1-x^{3/2}\right|^{-1}
\eeq
with $x$ as the solution of the equation
\beq
x^{1/2}{1-8x^{3/2}/7\over 1-x^{3/2}} = {0.64\over {\dot M}_{17}} ~ .
\eeq

Figure 2 plots the spin equilibrium and spin-up curves for accretion rates 
appropriate to GK Per in quiescence, $\dot M \sim$ a few $\times ~ 10^{16}$
g s$^{-1}$ (Bianchini \& Sabbadin 1983), and at maximum, $\dot M \sim$
$10^{19}$ g s $^{-1}$ (\S2).  The GK Per system should fall between the 
two curves if our magnetic disk model is reasonably accurate. 
Figure 2 assumes that the accretion disk and disk-magnetosphere 
interaction is in steady-state. The GK Per disk is, however, genuinely 
time-dependent, and a derivation of the white dwarf spin should include
the time-variability.  A complete theory of spin equilibrium for a 
time-dependent disk is beyond the scope of this paper.  We consider 
the time-averaged properties as a compromise.  For a recurrence time scale 
of $\sim 1000$ d and an outburst duration of $\sim 50$ d, the time-averaged 
accretion rate is $\sim 1-3\times 10^{18} ~ \rm g ~ s^{-1}$.  The observed 
spin equilibrium may correspond to an average accretion rate that is much 
smaller than the outburst accretion rate needed for the UV emission.  
If we select magnetic parameters appropriate for spin equilibrium at 
this accretion rate, disks in outburst spin up the star and have less 
magnetic heating than disks on the spin equilibrium line (see below).  
The temperature distribution then closely approximates a steady-state disk,
which does not explain the UV emission (Figure 1).

In addition to the coherent X-ray pulsation, GK Per also has optical
quasi-periodic oscillations (QPOs) with amplitudes of a few per cent.
The observed QPO period, $P_{QPO}$,  depends on state of the system.  
Patterson (1981) first identified $P_{QPO}\sim 380$ s during quiescence in 
1978; Mazeh et al. (1985) later estimated a QPO period $P_{QPO}\sim 360$ s
during the decline phase from the 1983 data. These two estimates are 
consistent within the observational errors. Mazeh et al. (1985) also found 
$P_{QPO} \sim$ 400 s on two nights in outburst.  

The QPO periods provide another constraint on the magnetized accretion 
disk model. This constraint is, however, highly model-dependent due 
to the lack of a unique model for QPOs. In the beat frequency model
(e.g. Lamb et al. 1985), blobs orbiting at a frequency slightly shorter 
than the stellar rotation period produce QPO at the inner edge of the disk.
The Keplerian period at $R=R_0$ is 
\beq
P_0=P_{QPO}P_*/(P_{QPO}+P_*).
\eeq
The observed QPO periods indicate $P_0 \approx$ 180 s and $R_0 \approx$
$4.6\times 10^9$ cm in quiescence, with $P_0 \approx$ 190 s and 
$R_0 \approx 4.8\times 10^9$ cm in outburst.  This behavior is opposite
to our expectations, because the inner radius should decrease when 
$\dot M$ is large (see equations (3-8) and (3-9)).  This change could 
indicate slight spin-up, although we cannot make any quantitative conclusions.
Nevertheless, Fig. 2 plots the curve for $P_{QPO}\sim 400s$ during outburst. 
Again using the beat frequency model, 
the $B_*-{\dot M}$ curve for a given $P_{QPO}$ is 
\beq
{\dot M}_{17}=3.34\times 10^2\left(1-y^{3/2}
\over y^{7/2}\right)B_{eff,7}^2
\eeq
where
\beq
y=\left(P_{QPO}\over P_{QPO}+P_*\right)^{2/3}
\eeq
which becomes $\sim 0.66$ for $P_*=351s$ and $P_{QPO}\sim 400s$.
The relation between the accretion rate and the magnetic field 
strength is then ${\dot M}_{17}=36.2B_{eff,7}^2$.

Patterson (1991) considered a QPO model where blobs {\it outside}
corotation produce periodic light variations and estimated
$R_{blob} \approx 9 \times 10^{10}$ cm in quiescence and
$R_{blob} \approx 3 \times 10^{10}$ cm in outburst.  Although
these large radii are inappropriate for the inner disk radius, 
their ratio is comparable to the radius variation expected in
a magnetic disk model when $\dot M$ increases during an eruption.
We speculate that non-axisymmetric disk features outside the 
corotation region could produce the observed QPOs.  X-ray
irradiation of local density enhancements in a non-steady disk could 
produce bright optical blobs that give rise to QPOs.  These density
enhancements move radially with time in disk instability models (e.g.,
Kim et al. 1992) and might produce the observed QPO period changes 
described above.  Detailed disk instability models with X-ray
irradiation would test this hypothesis.

Having isolated magnetic parameters appropriate for GK Per, 
we now consider the temperature distribution of a magnetically
truncated accretion disk derived in a self-consistent way from
the magnetic torque.  This torque has its largest affect on the
temperature distribution when the disk lies well above the spin
equilibrium line and the star is spinning down.  The torque can
modify the temperature distribution significantly on the
spin equilibrium line, but the disk quickly approaches the 
non-magnetic temperature distribution when $\dot M$ and $B_*$ act 
to spin up the star.  We thus do not consider magnetic heating on
the QPO curve in Figure 2.

Figure 3 shows temperature distributions for 10 models along
the spin equilibrium and spin-up curves of Figure 2.  Each pair
of models has the same $\dot M$ and $R_0$; the spin-up models have 
smaller field strengths and hence less magnetic heating.  In all 
cases, the spin equilibrium model has a lower temperature near $R_0$
and a higher temperature outside $R_0$ compared to the spin-up model.
The temperature gradient outside $R_0$ is fairly independent of the
magnetic heating and is shallower than $T \propto R^{-3/4}$ in all 
of the models.

Figure 4 plots UV spectra for several magnetic disk models along
the spin equilibrium and spin-up curves.  Although the temperature
distributions of these models are flatter than steady disks, the
model fluxes fit the data poorly.  The lowest $\dot M$ models
(1 and 2) have nearly the correct UV slope, but the UV fluxes
lie a factor of 10--20 below the observations.  These spectra are 
similar to the spectra of models DI and SS2 in Figure 1.  The predicted 
fluxes reach the correct luminosity for $\dot M \approx$
$10^{19} \rm ~ g ~s^{-1}$ (models 7 and 10).  These models have 
shallower UV slopes than steady non-magnetic disk models at the 
same accretion rate, but are too blue compared to observations.

\subsection{Other Magnetized Disk Models}

Our results indicate that magnetic heating models cannot explain the
UV spectrum of GK Per at maximum light.  Before we abandon this model
completely, we consider if other treatments of a magnetized disk have
a better chance of reproducing observations.  For instance, 
Mauche et al. (1990) introduced two additional parameters to model 
the magnetic pitch distribution on phenomenological grounds. 
Force-free models also treat the magnetic stress differently and
could lead to an ``improved'' temperature distribution compared to
our derivation. 

In the original Ghosh-Lamb model (1979ab), the wind-up of the azimuthal
field component is given by
\beq
\left|\partial B_{\phi}\over\partial t\right|_+=\gamma\left|\Omega_*-\Omega\right|
\left|B_{\phi}\right| .
\eeq
The magnetic field changes on a characteristic reconnection time scale 
$\tau_{d}\approx H/\xi\left|v_{A\phi}\right|$ where $\left|v_{A\phi}\right|$ 
is the Alfv{\`e}n velocity associated with $B_{\phi}$ and $\xi<1$ is a numerical
reconnection factor. The equilibrium field distribution is 
\beq
\left|B_{\phi}\over B_z\right|=\left|{\gamma\over \xi}{H\over v_{Az}}\left
(\Omega_*-\Omega\right)\right|
\eeq
where $v_{Az}$ is the Alfv{\`e}n velocity corresponding to $B_z$.
Following a procedure similar to \S3.2, the torque on the star is
$$
{N\over N_0}=1+{64\over 21}{(R_0/R_c)^{31/16}\over 1-(R_0/R_c)^{3/2}}\left[
{7\over 8}\left(R_0\over R_c\right)^{3/16}+{1\over 8}\left(R_0\over R_c\right)
^{-21/16}-1\right]\qquad\qquad\qquad\qquad
$$
\beq
-{64\over 21}{(R_0/R_c)^{31/16}\over 1-(R_0/R_c)^{3/2}}
\left[{7\over 8}\left(R_{out}\over R_c\right)^{3/16}+{1\over 8}\left(R_{out}\over
R_c\right)^{-21/16}-1\right].
\eeq
The disk truncation radius $R_0$ is set by eq. (3-9) with $A$ replaced by
\beq
A=[4.4\times 10^3]\left(\gamma\over \xi\alpha^{9/20}\right)B_{*,7}R_{*,9}^3
M_{*,1}^{-2/3}P_{*,2}^{-29/24}{\dot M}_{17}^{-23/40}
\eeq
where we have adopted the gas pressure and the Kramers' opacity for the disk 
structure (e.g. Frank et al. 1992). 
For $R_{out}/R_c\approx 20$, spin equilibrium is reached when 
$R_0/R_c=0.591$, which indicates that the disk extends further inward
from the corotation radius than our present model. 
The difference in the disk temperature distribution becomes most significant
outside the corotation radius. The ratio of the magnetic pressure $P_{B}$ 
to the gas pressure $P_g$ is $P_B/P_g\sim (\gamma/\xi)^2(R/R_c)^2$, so
magnetic dissipation dominates at $R\gg R_c$. In this region, 
\beq
{dL\over dR}\sim {\gamma\over \xi}{H\over v_{Az}}\Omega_*^2{B_c^2R_c^6\over R^4}
\propto {\gamma\over \xi}R^{-13/16}
\eeq
and
\beq
T_{eff}\propto R^{-29/64}\sim R^{-1/2}
\eeq
which is much flatter than the familiar $T_{eff}\propto R^{-3/4}$ temperature 
profile.  This result is particularly encouraging for GK Per as outlined
earlier.  We do not further explore this model, however, due to the arbitrary 
prescription for the azimuthal field amplification.  The dynamical structure 
of the outer region with $P_B>P_g$ is also largely unphysical (Wang 1987). 

An alternative model requires the magnetosphere to satisfy the force-free 
condition, i.e. $\nabla {\bf B}\propto {\bf B}$.  The magnetic field lines 
then continuously reconnect to prevent the growth of the magnetic stress
inside the magnetosphere.  Magnetic reconnection, which is the field 
loss mechanism, occurs on a time scale
$\sim\gamma\left|\Omega_*-\Omega\right|\approx\Omega_*$ for $\Omega_*>\Omega$ and
$\approx\Omega$ for $\Omega_*<\Omega$ (Aly \& Kuijpers 1990).
Eq. (3-6) then indicates that the magnetic pitch must lie below a constant 
of order unity.  We can thus describe this model by replacing $\gamma/\alpha$ 
with a constant $\gamma_{max}\sim O(1)$ and $\Omega$ in the denominator of 
eq. (3-6) with $\Omega_*$ for $R>R_c$ where $\Omega_*>\Omega$. 
The resulting torque expression produces an equilibrium spin at 
$R_0=0.950R_c$ (Wang 1995).  For $\gamma_{max}=\gamma/\alpha$, this
model is almost indistinguishable from our model in the region $R<R_c$. 
The disk temperature distribution for $R>R_c$ is also adequately approximated
by our model.

\section{X-ray Heating}

X-ray heating is our final alternative to produce a flat temperature 
gradient in GK Per's accretion disk.  In this mechanism, the disk
absorbs radiation from the base of the accretion column -- and perhaps 
the surrounding white dwarf photosphere -- which raises the local
blackbody temperature.  Wade (1988) and Smak (1989) show that this 
process is unimportant for most CVs, because the disks are small and 
radiation from the boundary layer has a large dilution factor in the 
disk plane.  However, the disk in GK Per is several times larger than 
a typical CV disk, and the accretion column more easily illuminates a 
disk than a boundary layer.  

To establish that this process is viable for GK Per, we consider disk
heating near maximum in outburst.  The accretion rate is then $\sim$
$10^{19}$ g s $^{-1}$ (e.g., Kim et al. 1992), which produces a
luminosity of $\sim$ 250 $L_{\odot}$ at the white dwarf's surface.
For simplicity, we assume the central star radiates this luminosity
uniformly in all directions and parameterize the height of the disk 
photosphere above the midplane as $H/R = 0.04 R^n$ (see Warner 1995; 
Kenyon \& Hartmann 1987).  A truncated disk with parameters appropriate 
for GK Per intercepts $\sim$ 2\% of the accretion luminosity for a 
standard disk with $n = 0.125$;  the fraction increases to $\sim$ 6\% 
for $n = 0.25$ and $\sim$ 19\% for $n$ = 0.5 (see Kenyon \& Hartmann 1987).  
If the disk radiates all of this energy in the continuum, the extra 
luminosity from X-ray heating is 5--50 $L_{\odot}$.  This extra 
luminosity is close to the total UV luminosity at maximum, 
$\sim$ 20 $L_{\odot}$ (Wu et al. 1989).

This calculation is overly simplistic, because it does not consider
the geometry of the accretion column or the details of X-ray absorption
in the disk atmosphere.  In a standard magnetic accretion geometry,
gas accreting onto the central star along magnetic field lines should
produce two rings of emission at latitudes, $\pm$b, measured from the
stellar equator (see Patterson 1994 and references therein).  
Standing shocks form above each ring at a height (e.g. Yi et al. 1992)
\beq
H_s=[2\times 10^8 {\rm cm}]s_{\circ}M_{*,1}^{23/14}B_{*,7}^{-4/7}
R_{*,9}^{-3/14}{\dot M}_{17}^{-5/7}
\eeq
where $B_{*,7}=B_*/10^7G$ and $s_{\circ}$ is a constant of order unity. 
This approximation is valid even when the magnetic axis is not aligned
with the rotational axis, although the rings are then only partially
filled with emission (K\"onigl 1991; Hellier 1993).  This height is small
compared to the stellar radius at high accretion rates, $\dot M >$ 
$10^{-18}$ g s$^{-1}$, and we can reasonably approximate the accretion
hot spots as a point displaced from the center of the white dwarf.
Our simple estimates for the accretion luminosity intercepted by the disk 
are then valid.  However, the height of the accretion shock increases as 
$\dot M$ decreases and becomes $H_s \sim R_*$ for $\dot M \le 10^{18}$ 
g s$^{-1}$.  We then cannot simply approximate accretion hot spots as
a point close to the white dwarf photosphere, which greatly complicates
our calculation of disk heating.

A direct calculation of X-ray processing by the disk is also complicated.
Previous results suggest that the disk is very optically thick to hard X-rays;
this radiation heats the disk to produce extra continuum and line emission
(Raymond 1993 and references therein).  Raymond's (1993) models for Sco X-1 
and other luminous X-ray binaries show that X-rays penetrate close to the 
disk photosphere at small disk radii, $R \sim 10^{8}$ to $10^9$ cm,  where 
electron scattering is the dominant opacity source.  These disk radii emit 
approximately as blackbodies.  Line emission is more important at large disk
radii, $R \sim 10^{10}$ to $10^{11}$ cm, where the X-ray optical depth 
exceeds unity well above the disk photosphere.  Prominent H~I Balmer and 
He I, He II emission continua also modify the blackbody spectrum of the 
outer disk in Raymond's (1993) models.  

The X-rays in GK Per are a factor of 10 or more weaker than expected from
the UV luminosity, so we might expect considerable EUV heating of the disk.
Proga et al. (1996) considered a red giant atmosphere illuminated by a hot
white dwarf with $T_{eff} = 2 \times 10^5$ K and showed that the atmosphere
re-emits most of the incident radiation as emission lines and a strong 
recombination continuum.  The illuminated atmosphere emits 25\%--50\% of 
the incident flux as continuum radiation for conditions less extreme than 
those in GK Per.  We thus expect a large contribution from H~I Balmer 
continuum radiation if the accretion column emits mostly in the EUV.  
This radiation would flatten the UV continuum considerably compared 
to optically thick disk models (see below).

To place better limits on the importance of X-ray heating in GK Per, we 
assume a point-like X-ray source illuminating a disk whose photospheric
height varies as $R^n$.  The X-ray heating flux at a disk radius R is 
roughly
\beq
F_x={f\eta G M_*{\dot M}\over R^2R_*}.
\eeq
where $\eta\le 1$ is the uncertainty of the X-ray efficiency at the 
central source (e.g. Frank et al. 1992) and $f$ is the X-ray absorption 
efficiency that takes into account the X-ray albedo $\sim 0.3$ and 
the X-ray heating geometry.  
The factor $f$ also includes the radial variation of the projected 
surface area of the disk and thus slowly varies as a function of $R$ 
(see below). This heating source dominates locally generated accretion 
energy for disk radii,
\beq
R>R_x=[1.5\times 10^{11}{ \rm cm}]R_{*,9}\left(f\eta\over 10^{-2} \right)^{-1} ~ .
\eeq
Thus heating begins to dominate accretion near the outer edge of
the disk in GK Per, $R_{out} \approx 1.5 \times 10^{11}$ cm, for
$f\eta \sim 0.1$.  The temperature profile is then intermediate
between the $T \propto R^{-3/4}$ of a steady disk and the 
$T \propto R^{-1/2}$ limit of a pure reprocessing disk (see 
Smak 1989, Kenyon \& Hartmann 1987, and references therein).  
For simplicity, we model $F_x$ as a local heating term and compute 
the radial temperature gradient in the blackbody approximation:
\beq
T_{eff}=\left[{1\over\sigma R}\left(F_{vis}+F_{x}\right)\right]^{1/4} ~ .
\eeq

Several factors complicate the direct calculation of $f$ needed to derive a 
self-consistent temperature distribution for an X-ray heated disk.  In addition
to the geometry of the accretion column, the white dwarf shields some portions
of the disk from hard X-rays when the magnetic and rotational axes are
not aligned.  If the angle between these two axes is $\phi$, the innermost
disk radius exposed to X-rays is approximately
\beq
R_I=R_{*}\left({1\over \sin\phi}-1\right).
\eeq
In GK Per, the region near the truncation radius at $R\sim R_I\sim 10R_*$ 
is shielded for $\phi \approx 6^{\circ}$, which is comparable to the
misalignment angles measured in other CVs (see Hellier 1993 and references
therein). 

We compute $f$ for $R>R_I$ assuming the disk photosphere absorbs and
reradiates incident X-rays at $T_{eff}$ from equation (4-11).  We adopt
an X-ray albedo of 0.3 and the Shakura-Sunyaev variation of disk thickness 
with radius, $H/R$ (Frank et al. 1992; Warner 1995) and estimate
\beq
f\approx 2.1\times10^{-2}\left(\alpha\over 0.3\right)^{-1/10}
M_{*,1}^{-3/8}{\dot M}_{17}^{3/20}R_{10}^{1/8}
\eeq
for $H/R \ll$ 1, where $R_{10}=R/10^{10}$ cm.
For GK Per specifically,
\beq
f\approx 2.2\times 10^{-2}{\dot M}_{17}^{3/20}R_{10}^{1/8}.
\eeq
We note that the dependence on $\alpha$ in eq. (4-6) is very weak.
Given the uncertainties, we parametrize the X-ray heating as
\beq
f=f_{10}R_{10}^n
\eeq
where $f_{10}$ denotes $f=2\times 10^{-2}{\dot M}_{17}^{3/20}$ at 
$R=10^{10}$ cm. This expression recovers equation (4-7) for $n=1/8$.

Previous studies of disk heating show that `flat' disks with 
$n \le$ 1/8 have temperature distributions similar to a steady
disk with $T \propto R^{-3/4}$ (Kenyon \& Hartmann 1987).  Figure
5 shows temperature distributions for four combinations of 
$({\dot M}, B_{eff})$ that lie between the spin equilibrium line 
and the spin up line at each ${\dot M}$ in Fig. 2. 
In each panel, the flattest temperature distribution corresponds to 
$n=1.2$, the middle one to $n=0.6$, and the steepest one to $n=0.3$.

Figure 6 compares spectra of disks with a low X-ray heating rate ($n=0.3$)
and disks without X-ray heating.  The extra luminosity of the heated 
disks is considerable, even when the disk height increases slowly with 
radius as in standard hydrostatic disk models.  However, models with the 
correct UV slope are underluminous compared to observations.  The
UV slopes of more luminous disks are too blue compared to the data.
The agreement between the models and the data increases with increasing $n$.  
Figure 7 compares spectra for two other model temperature distributions --
model 27 and model 28 -- with observations of GK Per.  Both models
match the data over 1750--3250~\AA~and are a factor of $\sim$ 2 above 
the data at shorter wavelengths.  Both models predict V magnitudes 
within $\pm$0.3 mag of those observed near maximum.
Less extreme heating models also predict the observed V but fail to
agree with the UV data over as large a wavelength range. 

Although we find some agreement between models and observations with
X-ray heating, the resulting variation of disk height with radius is
severe.  A disk with $n = 1.2$ is extremely concave and physically 
unrealistic.  Previous calculations of disk heating indicate that 
the disk probably maintains hydrostatic equilibrium with 
$n \le$ 0.3--0.5 (e.g., Kenyon \& Hartmann 1987; Raymond 1993 and
references therein) and cannot reach $n \ge$ 0.5.  Even if an $n = 1.2$
disk can be achieved in nature, the outer rim of this disk occults most
of the inner disk for modest inclinations, $ i \approx 45^{\circ}$.
Thus, this disk model is probably inappropriate for GK Per.

Finally, we consider a simple optically thin disk model.  We adopt 
the X-ray heating model described above for $n = 0.3$, assume X-ray 
photons ionize material above the disk photosphere, and calculate the 
recombination continuum for an LTE slab of hydrogen with an optical depth
at 3000~\AA, $\tau_{3000} \approx$ 0.5. 
We adopt a slab temperature of 15,000 K,
which results in a total UV luminosity, 5 $L_{\odot}$, comparable to
the accretion energy intercepted by the disk.  Although these parameters
are not `best' for an accretion disk corona in GK Per, this calculation 
provides a first approximation to optically thin emission in the heating 
model.  Figure 8 compares spectra of spin equilibrium model 8 with 
and without a recombination continuum from our accretion disk corona.  
The recombination continuum flattens the UV spectrum considerably, 
although the predicted spectrum falls below the observations at all 
UV wavelengths.  This result is sensitive to our adopted parameters.  
Hotter coronae produce very blue spectra; smaller optical depths emit 
too little UV flux.  A better calculation, similar to Raymond's (1993) 
treatment, is needed to show if this hypothesis can explain the spectrum 
in detail.

\section{Summary and Discussion}

We have investigated several accretion disk models to account for
the observed UV spectrum of GK Per.  This spectrum indicates a 
large inner disk radius and an unusually flat temperature distribution. 
These requirements are not satisfied by any non-magnetic steady disk 
model or by existing disk instability models that are otherwise successful
in explaining the recurrence time scales of GK Per's occasional 
dwarf nova outbursts. 

Magnetic disk models also fail to explain the UV spectrum.  We used 
data for QPOs and the spin-down rate to constrain the inner radius
of an equilibrium system and found that $\dot M$ and $B_*$ are
confined to a narrow band with $36 \le {\dot M}_{17}/B_{eff,7}^2 \le 57$.
Magnetic torque significantly affects the disk temperature distribution 
at the upper end of this range and produces shallower temperature laws
than predicted for standard steady disks.  Our model temperature
distributions are, however, steeper than needed to explain the 
observed UV spectrum.

X-ray heating models are a promising mechanism to produce a flat
disk temperature gradient in GK Per.  Optically thick models that
absorb and reprocess X-rays close to the disk photosphere require
significant curvature in the disk to reproduce the UV spectrum.
These disks are physically unrealistic.  Optically thin models can
explain the UV spectrum if Balmer continuum radiation dominates the
emitted spectrum of an accretion disk corona.  Our example in Fig. 8
represents a favorable case where the disk+corona intercepts 5\%--10\% 
of the accretion luminosity and remains cool enough to emit much of 
this energy at 2000--3000~\AA.  More detailed calculations are needed 
to test the viability of this solution.

The heating model has several consequences that can be directly tested 
by observation.  If much of the 2000--3000~\AA~continuum is H~I Balmer
emission, this flux should show large amplitude variations phased with
the X-ray light curve.  Patterson (1991) detected stable variations at
U with the appropriate period.  The amplitude of this variation is small,
$\sim$ 0.004 mag, compared to the X-ray amplitude.  Observations of
emission shortward of the Balmer jump would provide a better test of
the heating model.  The optical emission line spectrum provides another
test.  Measurements of the H~I line fluxes through an outburst directly 
constrain the flux from the Balmer continuum; detection of periodic 
variations in line flux phased with the X-ray period places additional 
limits on the heating model.

Even if X-ray heating does not explain the UV spectrum directly, this
heating is an important contribution to the energy balance of the
disk itself.  Our estimates in \S4 indicate that the disk intercepts
$\sim$ 5\% of the accretion energy radiated by material impacting the
surface of the white dwarf.  This energy is a significant fraction,
$\sim$ 10\%--20\%, of the disk luminosity in outburst and probably 
changes the vertical temperature structure of the disk.  This heating 
could be larger during quiescence when the height of the accretion 
shock above the white dwarf increases (cf. eq. (4-1)).  The vertical
temperature structure is an important feature of disk instability models 
that trigger outbursts when disk material is too cold to radiate viscous 
energy generated locally (see Kim et al.  1992 and references therein). 
Thus, X-ray heating might modify the properties of dwarf nova-like 
behavior in this system.

The accretion rates required for the observed UV emission at maximum
exceed $10^{19} ~\rm g~s^{-1}$ and are incompatible with the
observed X-ray luminosity. The observed X-ray luminosity ($L_x$) 
implies ${\dot M}\sim L_x R_*/G M_*< 10^{18} g/s$ (Ishida et al. 1992, 
Norton et al. 1988, Yi et al. 1992, Yi 1994, Yi \& Vishniac 1994) 
in the X-ray emitting magnetic accretion column.  The X-ray luminosity
remains more than an order of magnitude higher than the optical/UV
luminosity in quiescence, which further contradicts the results of
disk instability calculations (Yi et al. 1992, Kim et al. 1992).

In a magnetically channeled accretion column (Frank et al. 1992), 
the radially free-falling material shocks above the white dwarf
photosphere.  Highly ionized material in the postshock gas emits
X-ray through bremsstrahlung.  Due to highly anisotropic (Compton) 
scattering geometry, most photons travel radially outward along the 
column axis and exert a radiation drag on the infalling material.
As a result, the X-ray emission temperature can be substantially
lower than the expected postshock temperature $\sim 5\times 10^8 K$.
This effect becomes significant for high ${\dot M}$ and
small column cross-sectional area (Yi 1994, Yi \& Vishniac 1994).
For typical parameters derived from the UV spectrum and spin, 
the ratio of the column cross-sectional area to white dwarf surface 
area is $\sim 0.06s_o ({\dot M}/10^{18} g/s)^{2/7}$, where $s_o\le 1$ 
is a constant (Yi 1994).  The radiation-modified postshock (X-ray emission) 
temperature falls below $10^8$K when ${\dot M}\sim 4\times 10^{18}\rm~g~s^{-1}$ 
for $s_o\sim 0.5$ (Yi 1994, Yi\& Vishniac 1994).  The X-ray temperature
continues to fall as $\dot M$ increases; the X-ray luminosity is then
much less than the total accretion luminosity.  For example, the
X-ray luminosity for E $>$ 2 keV at ${\dot M}\sim 4\times 10^{18} g/s$ 
implies an apparent ${\dot M} \sim 6\times 10^{17} \rm ~g~s^{-1}$.
The predicted X-ray emission spectra (after photoelectric absorption)
tend to show little indication of this effect (Yi \& Vishniac 1994).
We conclude that the mass accretion rates in UV and optical can be 
reconciled with the observed X-ray luminosity. This effect requires
${\dot M}>10^{18} g/s$ which is consistent with the observed UV emission.

Finally, combining spectral information and improved spin torque 
measurements could convincingly constrain physical parameters which 
are largely unknown. Precise measurements of spin-up (-down) torque 
during the rise (decline) phase of outburst could provide especially
valuable information on the magnetized accretion disk model.
More reliable data on QPO during quiescence and outburst could
further constrain the magnetized accretion disk model.

\acknowledgments
We thank D. Proga, J. Raymond, and J. Craig Wheeler for helpful advice
and discussions. I. Y. acknowledges support from SUAM Foundation.
S. K. acknowledges support from NASA grant NAG5-1709.

\clearpage

\begin{figure}[t]
\centerline{\psfig{figure=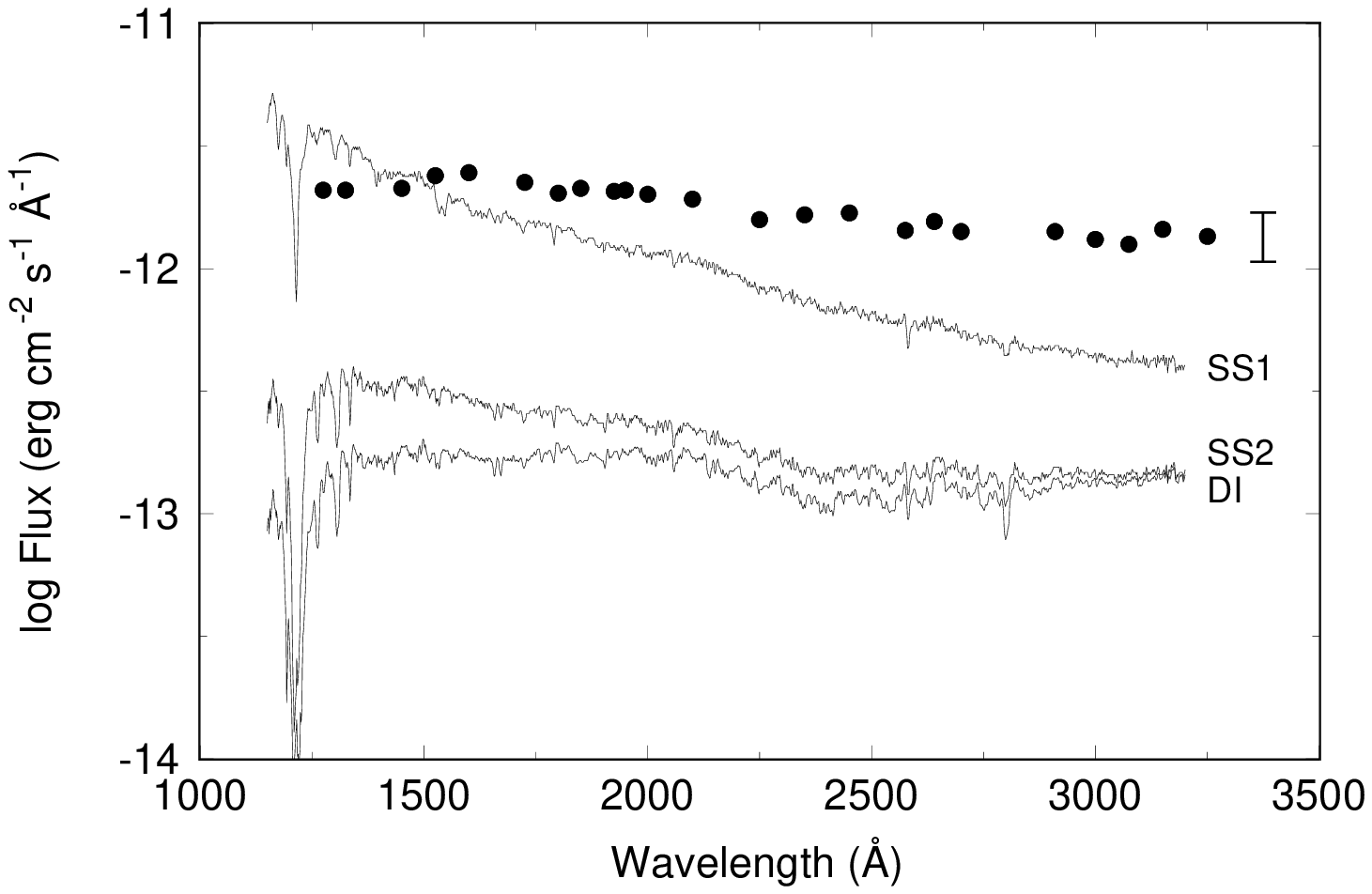,width=14.0cm,height=14.0cm}}
\caption[]
{Ultraviolet spectra of the two steady disk models and the Kim et al. (1992)
disk instability model (see text). 
Filled circles plot dereddened {\it IUE} fluxes for spectra acquired
at maximum during an outburst (SWP 13497 and LWR 10143 on 1981 March 15;
Wu et al. 1989). For inclinations $i>45^{\circ}$, these
models cannot fit the spectral slope and the flux level simultaneously.
}
\end{figure}

\begin{figure}[t]
\centerline{\psfig{figure=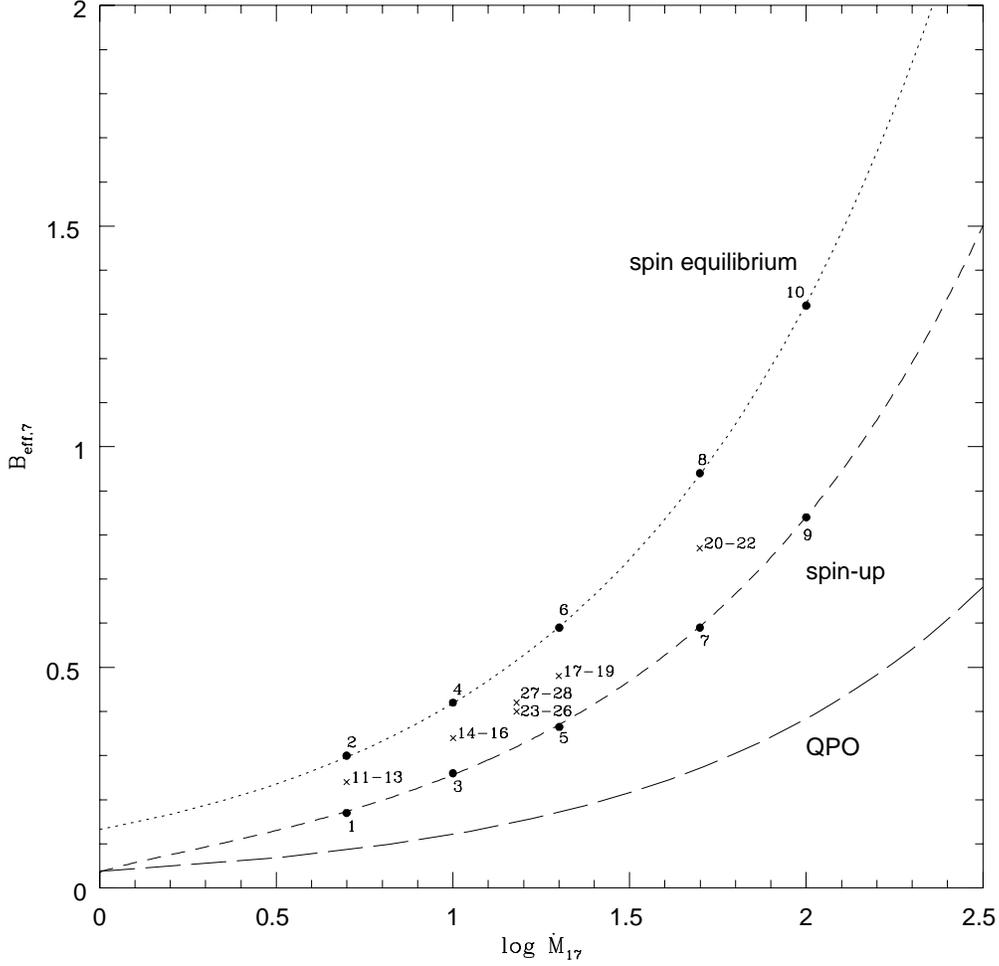,width=14.0cm,height=14.0cm}}
\caption[]
{Magnetic field ($B_{eff,7}=\sqrt{\gamma/\alpha}B_*/10^7G)$ $-$
mass accretion rate (${\dot M}_{17}={\dot M}/10^{17}g/s$) parameter
space. The long dashed line corresponds to the observed QPO period of
$P_{QPO}=400s$ in the beat frequency model. The short dashed line 
corresponds to the spin-up torque derived in Ishida et al. (1992). 
The dotted line corresponds to spin equilibrium. The parameter space 
above the spin equilibrium line -- the spin-down region -- is ruled 
out from the observed spin periods. Ten models marked by numbers are 
considered for the standard magnetized accretion disk model. 
The temperature distributions corresponding to these models are shown in 
Figure 3. Models numbered 11-28 are X-ray heated models shown in various
other figures.}
\end{figure}

\begin{figure}[t]
\centerline{\psfig{figure=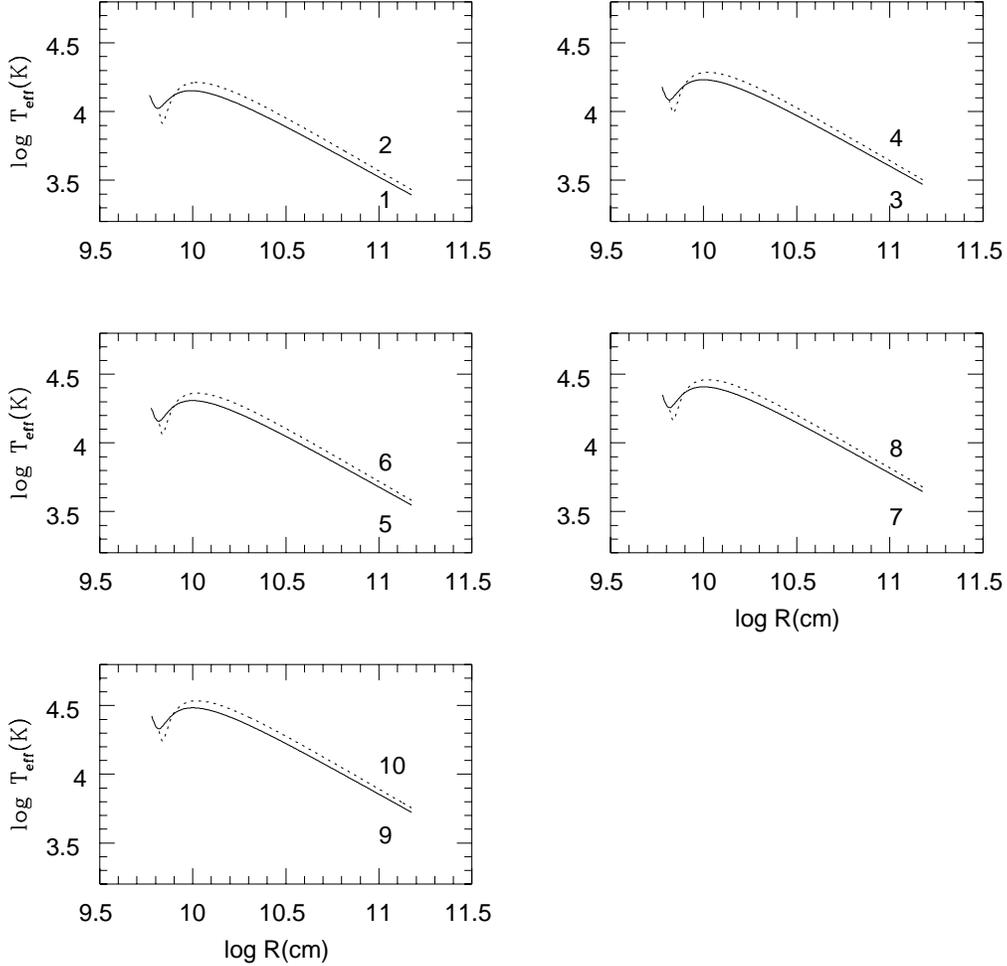,width=14.0cm,height=14.0cm}}
\caption[]
{The temperature distributions for the ten models in Figure 1.
The disk temperature distribution is set by the mass accretion 
rate and the magnetic torque heating. In each panel, the solid line 
corresponds to spin-up and the dotted line to spin equilibrium. 
Higher magnetic fields for spin equilibrium raise disk temperature 
through magnetic torque heating and shifts the disk temperature maximum 
to larger radii.} 
\end{figure}

\begin{figure}[t]
\centerline{\psfig{figure=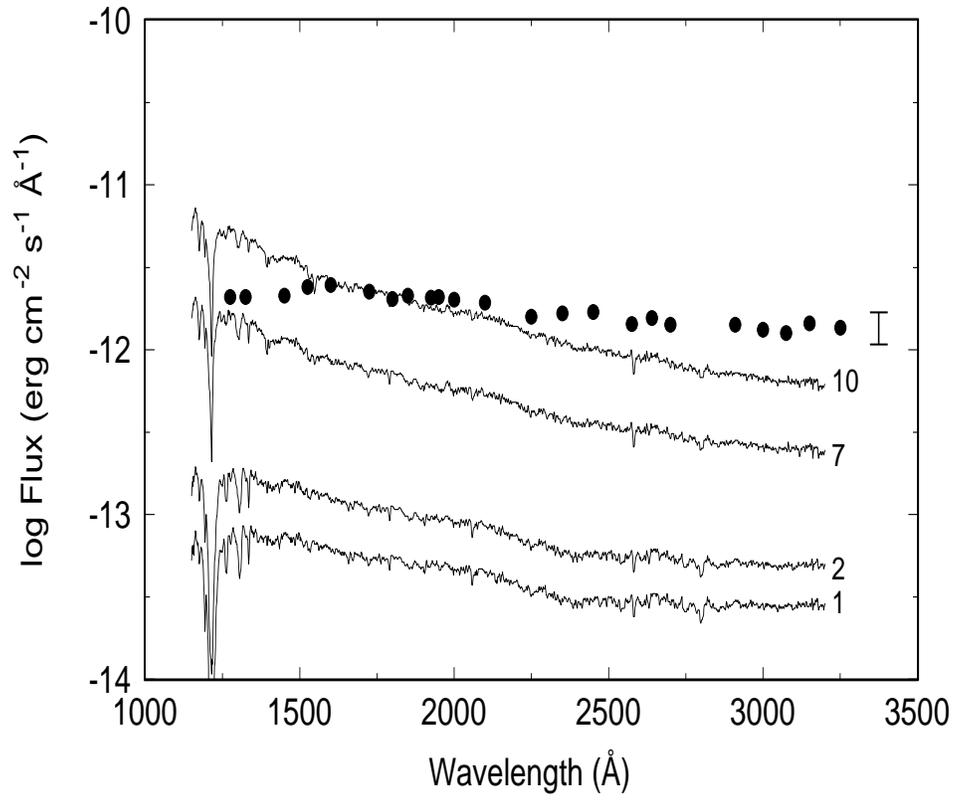,width=14.0cm,height=14.0cm}}
\caption[]
{UV spectra of some of the spin-equilibrium and spin-up models shown 
in Figure 2.
}
\end{figure}

\begin{figure}[t]
\centerline{\psfig{figure=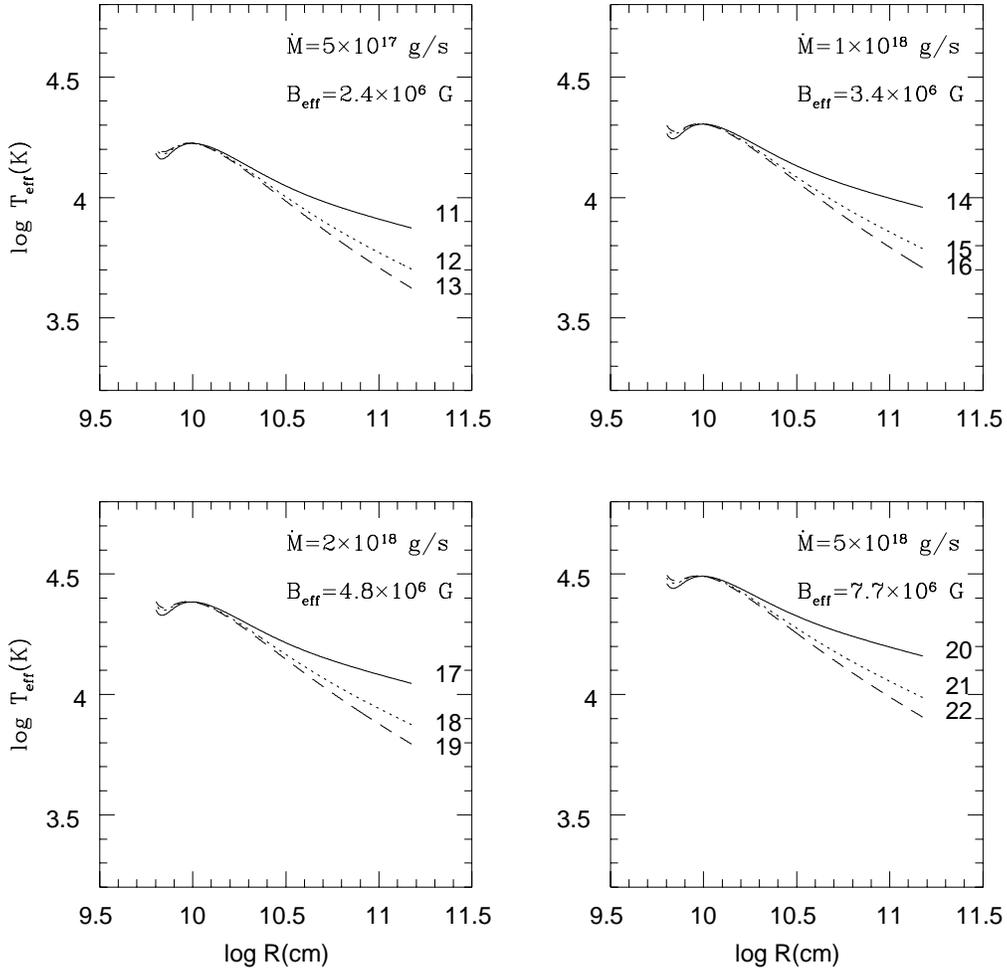,width=14.0cm,height=14.0cm}}
\caption[]
{Disk temperature distributions for X-ray heating models. The models
are all in slight spin-up allowed by the existing data. The parameters
are shown on each panel. In each panel, three different X-ray heating
models are shown (solid line: $n=1.2$, dotted line: $n=0.6$, dashed
line: $n=0.3$). The locations of the models in Figure 2 are marked by the
model numbers 11-22.}
\end{figure}

\begin{figure}[t]
\centerline{\psfig{figure=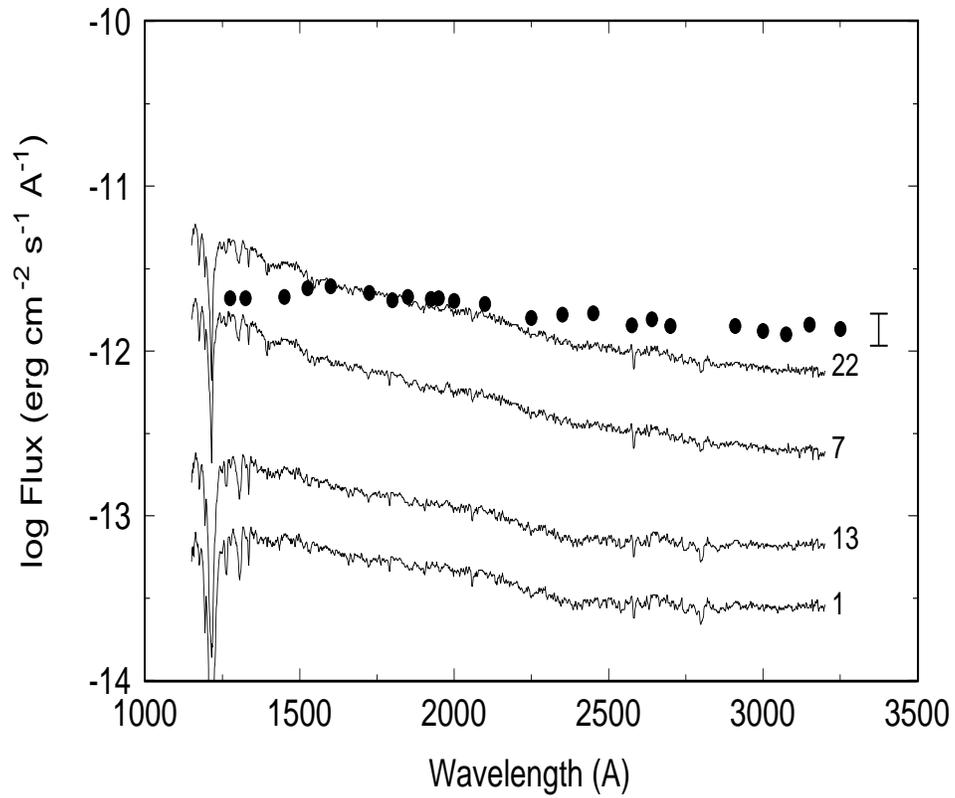,width=14.0cm,height=14.0cm}}
\caption[]
{Comparison of spectra of disks with low X-ray heating ($n=0.3$) and
disks without X-ray heating. The model numbers are identical as those
in Figure 2.
}
\end{figure}

\begin{figure}[t]
\centerline{\psfig{figure=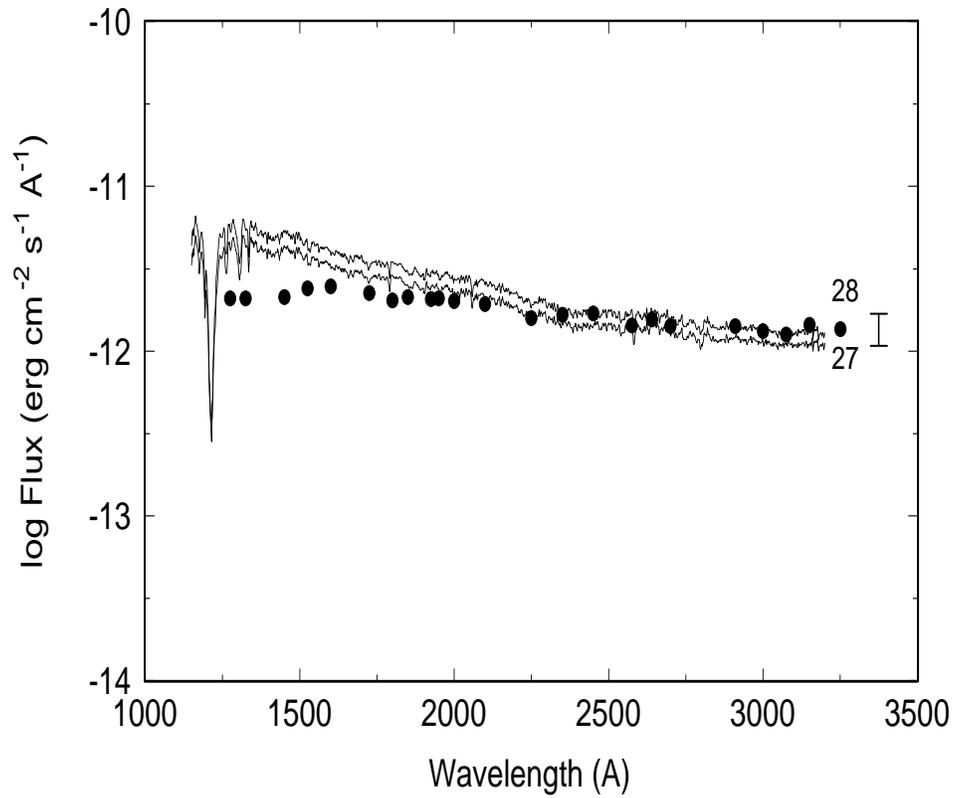,width=14.0cm,height=14.0cm}}
\caption[]
{UV spectra of two high X-ray heating models 27 ($n=1.3$) and 28 ($n=1.4$).
These good fits are due to the very flat temperature distributions resulting 
from a large X-ray heating rate as described in the text.
}
\end{figure}

\begin{figure}[t]
\centerline{\psfig{figure=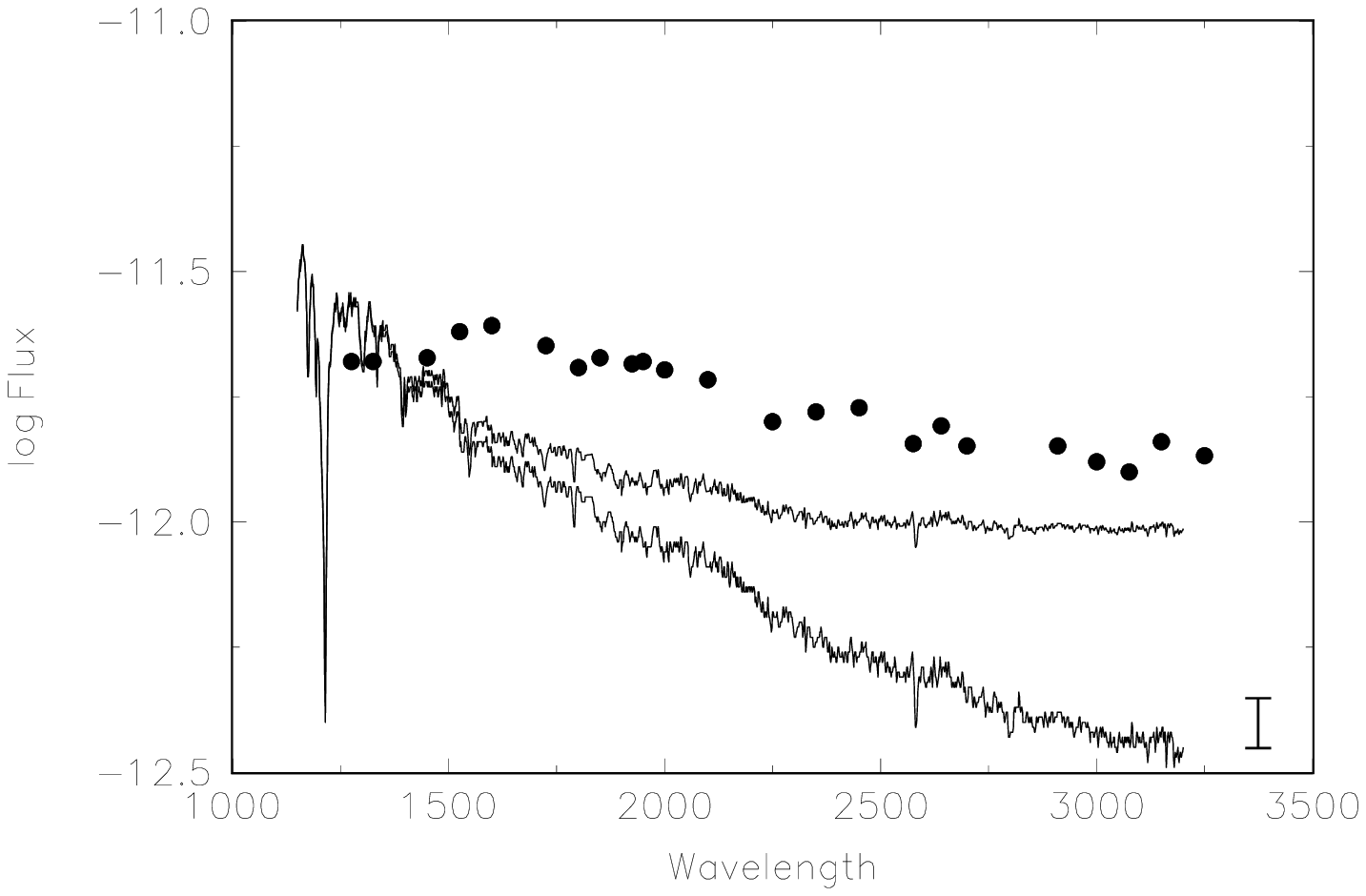,width=14.0cm,height=14.0cm}}
\caption[]
{Comparison of spectra of the spin equilibrium model 8 ($n=0.3$) with and
without additional continuum from a simple accretion disk corona.  
Models with a recombination continuum fit the data but are uncertain as
described in the text.
}
\end{figure}


\begin{references}

Aly, J. J. \& Kuijpers, J. 1990, A\&A, 227, 473

Angelini, L. \& Verbunt, F. 1989, MNRAS, 238, 697

Bianchini, A. \& Sabbadin, F. 1983, A\&A, 54, 393

Bianchini, A., Sabbadin, F., \& Dalmeri, I. 1986, A\&A, 160, 367

Campbell, C. G. 1992, Geophys. Astrophys. Fluid Dyn., 63, 179

Cannizzo, J. K. \& Kenyon, S. J. 1986, ApJ, 309, L43 

Cannizzo, J. K. \& Mattei, J. A. 1992, ApJ, 401, 642

Crampton, D., Cowley, A. P., \& Fisher, W. 1986, ApJ, 300, 788

Frank, J.,  King, A. R., \& Raine, D. 1992, Accretion Power in Astrophysics
(Cambridge: Cambridge Univ. Press)

Ghosh, P. \& Lamb, F. K. 1979a, ApJ, 232, 259

Ghosh, P. \& Lamb, F. K. 1979b, ApJ, 234, 296

Hellier, C. 1993, PASP, 105, 966

Ishida, M. et al. 1992, MNRAS, 254, 647

Kenyon, S. J., \& Hartmann, L. 1987, ApJ, 323, 714

Kenyon, S. J., Yi, I., \& Hartmann, L. 1996, ApJ, 462, 439

Kim, S.-W., Wheeler, J. C., \& Mineshige, S. 1992, ApJ, 384, 269

K\"onigl, A. 1991, ApJ, 370, L39

la Dous, C. 1991, A\&A, 252, 100

Lamb, F. K., Shibazaki, N., Alpar, M. A., \& Shaham, J. 1985, Nature, 
317, 681

Li, J. 1996, ApJ, 456, 696

Lovelace, R. V. E., Romanova, M. M., \& Bisnovatyi-Kogan, G. S. 1995,
MNRAS, in press 

Mauche, C. W., Miller, G. S., Raymond, J. C., \& Lamb, F. K. 1990, in
Accretion-Powered Compact Binaries (Cambridge: Cambridge Univ. Press), 195

Mazeh, T., Tal, Y., Shaviv, G., Bruch, A., \& Budell, R. 1985, A\&A, 149,
470

Nauenberg, M. 1972, ApJ, 175, 417

Norton, A. J. \& Watson, M. G. 1989, MNRAS, 237, 715

Norton, A. J., Watson, M. G., \& King, A. R. 1988, MNRAS, 231, 783

Patterson, J. 1981, ApJS, 45, 517

Patterson, J. 1991, PASP, 103, 1149

Patterson, J. 1994, PASP, 106, 209

Proga, D., Kenyon, S. J., Raymond, J. C., \& Miko{\l}ajewska, J. 1996,
ApJ, in press (10 Nov)

Raymond, J. C. 1993, ApJ, 412, 267

Ritter, H. 1985, A\&A, 148, 207

Shu, F. H. 1992, The Physics of Astrophysics (Vol. 2), Gas Dynamics
(Mill Valley: University Science Books)

Smak, J. 1984, PASP, 96, 5

Smak, J. 1989, Acta Astr., 39, 201

Smak, J. 1991, Acta Astr., 41, 269

Smak, J. 1993, Acta Astr., 43, 101

Wade, R. A. 1988, ApJ, 335, 394

Wang, Y. 1987, A\&A, 183, 257

Wang, Y. 1995, ApJ, 449, L153

Warner, B. 1985, in Cataclysmic Variables and Low-Mass X-ray Binaries,
ed. D. Q. Lamb \& J. Patterson (Dordrecht: Reidel), 269

Warner, B. 1995, {\it Cataclysmic Variable Stars}, Cambridge University 
Press, Cambridge

Watson, M. G., King, A. R., \& Osborne, J. 1985, MNRAS, 212, 917

Wu, C.-C. et al. 1989, ApJ, 339, 443

Yi, I. 1994, ApJ, 422, 289

Yi, I. 1995, ApJ, 442, 768

Yi, I. \& Vishniac, E. T. 1994, ApJ, 435, 829

Yi, I., Kim, S.-W., Vishniac, E. T., \& Wheeler, J. C. 1992, ApJ, 391, L25

\end{references}
\end{document}